# Ferroelectric materials and their applications in green catalysis

Weitong Ding,[a†] Xiao Tang,[b†] Wei Li,[c] Liangzhi Kou,[d*] Lei Liu[a,c,e*]

[a] Institute of Process Engineering, Chinese Academy of Sciences, Beijing 100049, China

[b] College of Science, Nanjing Forestry University, Nanjing 210037, China

[c] School of Chemistry and Chemical Engineering, Wuhan Textile University, Wuhan, 430200, China

[d] School of Mechanical, Medical and Process Engineering, Queensland University of Technology, Gardens Point Campus, QLD, 4001, Brisbane, Australia

[e] Dalian National Laboratory for Clean Energy, Dalian 116023, China

† These authors contributed equally to the work

Corresponding authors:

Liangzhi Kou, liangzhi.kou@qut.edu.au

Lei Liu, 2021047@wtu.edu.cn; liulei3039@gmail.com

**Abstract**: The demand for renewable and environmentally friendly energy source has attracted extensive research on high performance catalysts. Ferroelectrics which are a class of materials with a switchable polarization are the unique and promising catalyst candidates due to the significant effects of polarization on surfaces' physical and chemical properties. The band bending at the ferroelectric/semiconductor interface induced by the polarization flip promotes the charge separation and transfer, thereby enhancing the photocatalytic performance. More importantly, the reactants can be selectively adsorbed on the surface of the ferroelectric materials depending on the polarization direction, which can effectively lift the basic limitations as imposed by Sabatier's principle on catalytic activity. This review summarizes the latest developments of ferroelectric materials, and introduces the ferroelectric-related catalytic application. The possible research directions of 2D ferroelectric materials in chemical catalysis is discussed at the end. The review is expected to inspire the extensive research interests from physical, chemical and material science communities.

**Keywords**: ferroelectric materials; gas adsorption; catalysis;

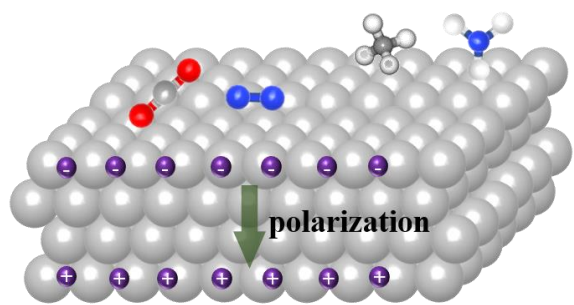

**TOC**: Traditional and 2D ferroelectric materials have been summarized, together with their applications in catalysis.

# 1. Introduction

With rapid development of science and technology, more and more energies are needed to meet the massive consumptions for human life and activities, however, the increasing consumption of fossil fuels has led to serious environmental pollutions. Hence, the catalytic reduction technology has attracted extensive attention because of its ability to produce renewable chemical fuels,[1-5] of which looking for catalysts with high-efficiency and selectivity is one of the key tasks. Recently, ferroelectric materials are considered to be promising candidates for the high-performance chemical reactions, since they are expected to overcome the limitations of the Sabatier principle. For example, ferroelectric $BiFeO_3$ can reduce the charge recombination rate of $BiVO_4$ anode from 17 $s^{-1}$ to 0.6 $s^{-1}$, and increase the oxygen evolution rate of photoelectrochemical (PEC) devices by about 4.4 times.[6] This enhancement effect is mainly attributed to the spontaneous polarization of ferroelectric materials. The generated built-in electric field inside the ferroelectric materials is beneficial to separate photogenerated electrons and holes. The charge imbalance induced by the field will be compensated by accumulating an equal amount of opposite charges on the surface by simple electronic reconstruction. Alternatively, electrons or holes are transferred to the opposite surfaces to compensate the charge imbalance,[7-9] by atomic reconstruction, namely adsorption or desorption of certain atoms on the surface of the materials. In addition, the different electrostatic potential and electron distribution of ferroelectric materials will also lead to different surface chemical activity and redox reaction capabilities. The chemical properties of the catalyst surfaces can be therefore controlled via ferroelectric switching, to improve the catalytic efficiency. Therefore, the application of ferroelectric materials in catalysts provides new opportunities for new catalysts with high performance.

With the advancement of thin film technology and nanoscale characterization methods, two-dimensional (2D) materials have triggered a new technological revolution. 2D materials with atomic thickness have a wide range of unique electronic,[10, 11] optical,[12, 13] mechanical[14] and thermal[15] properties, which do not exist in bulk counterparts[16]. It provides new ideas for the research of 2D ferroelectric materials. Generally, 2D materials consist of one or several atomic layers, which makes their ferroelectric properties significantly different from these of traditional bulk counterparts due to inherent size and surface effects. The depolarization field is significantly enhanced at 2D limits. Meanwhile, due to the reduction in the size of the material perpendicular to the plane, the surface of the 2D ferroelectric material has a richer morphology and active chemical environment, and its microstructure is easier to control. For example, polar nano-regions (PNR) in $SrTiO_3$ thin films with a thickness of less than 3 unit cells (3-UC) can be easily polarized. [17] The research of Kolpak and Wang showed that the special surface of 2D ferroelectric materials can effectively compensate the depolarization field.[18, 19] Moreover, van der Waals-type 2D ferroelectric materials have good durability against large strains. Unlike the surface of other 2D ferroelectrics, van der Waals-type 2D materials have a more stable surface structure, which may help maintain ferroelectric polarization,[20] as in $MoS_2$ and $In_2Se_3$.[21, 22] Up to date, a series of 2D ferroelectric materials have been theoretically predicted and verified in experiments, such as MXenes, group IV monosulfur compounds, $III_2$-$VI_3$ compounds, transition metal dichalcogenides, and transition metal phosphochalcogenides.

This review will start from a brief introduction of traditional ferroelectric materials, followed by current experimental synthesis and theoretical prediction of 2D ferroelectric materials, and then their applications in catalysis.

## 2. Traditional ferroelectric materials

Ferroelectric materials refer to the materials that have two or more spontaneous polarization directions within a certain temperature range, and the spontaneous polarization can be flipped under an external electric field.[23, 24] The essence of spontaneous polarization is the ion offset caused by the asymmetry of the crystal structure. Generally, the polarization of the materials depends on the external electric fields. Under the external electric field, the ion offset could be changed, which is called polarization inversion (**Figure 1a**). Therefore, ferroelectric materials show several important characteristics, such as dielectric, piezoelectricity, pyroelectricity, ferroelectricity, electro-optic effect, and acousto-optic effect, which could be used to fabricate ferroelectric memory, infrared detectors, and capacitive devices.[25] Traditional ferroelectric materials can be divided into five types according to their own structures: perovskite, pyrochlore, tungsten bronze, lithium niobate, and bismuth-containing layered structures. Among them, the ferroelectric material with the perovskite structure is the most important and the most studied one. The following contents will briefly introduce this type of structure, by taking the $BaTiO_3$ material as a representative example (**Figure 1b**).

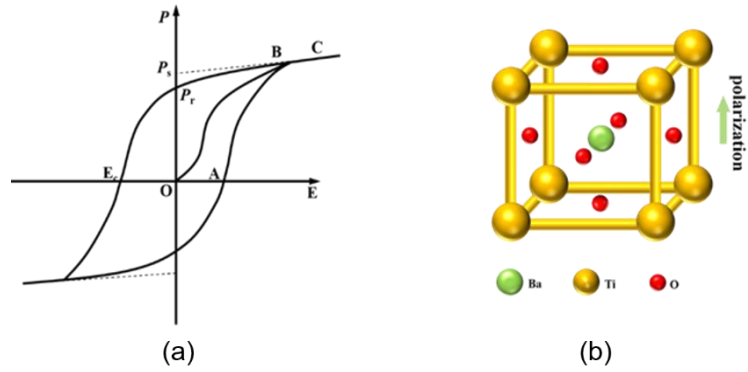

**Figure 1**. (a) Ferroelectric polarization curve; (b) Schematic diagram of atomic structures of perovskite ferroelectric materials.

With the change of temperature, $BaTiO_3$ undergoes a phase transition: the cubic paraelectric phase (~120 °C) to the tetragonal phase (~5 °C), and to the orthogonal phase (~-80 °C) trigonal phase.[26-29] At room temperature, the ferroelectric polarization is robust along the *c*-axis. When the temperature is higher than the Curie temperature (120 °C), $BaTiO_3$ is a cubic crystal system without ferroelectric features. The thermal motion of Ba is large enough to destroy the directional displacement effect of the internal electric field, and Ba locates in the center of the octahedron with the same probability of approaching the surrounding six O, thus, it does not have spontaneous polarization. $BaTiO_3$ becomes a tetragonal structure below 120 °C, and Ti shift from the original equilibrium position along the (001) direction, while the Ba is still located in the center of the tetrahedron with reduced thermal energy. Some Ti with lower thermal energy cannot overcome the internal electric field between Ti-O, and move toward a certain O (such as O in the *c*-axis direction), resulting in spontaneous polarization. When the crystal exhibits spontaneous polarization, a layer of positive and negative charges will be generated at both ends of the crystal, and an electric field opposite to the polarization direction is generated between the positive and negative charges, which is called depolarization. The electric field, which exists in the crystal, will increase the internal electrostatic energy, and render the structure unstable. In order to eliminate the effect of the built-in electric field, multiple regions with different polarization orientation are usually formed in the materials, which are the so-called ferroelectric domains. Moreover, the internal electric field can be offset by the reconstruction of electrons or atoms, leading to the different catalytic ability in the opposite surfaces of ferroelectric materials.

The size and fatigue effects of ferroelectric materials are important reasons to restrict their applications in electronic devices and chemical catalysis, where stable polarization can exist in ferroelectric thin films with the critical thickness of about tens of nanometers.[30] Ferroelectricity of traditional perovskite oxides will be significantly weakened or even disappear when it is thinner than several unit cells, such as 1.2 nm for $PbTiO_3$, 2.4 nm for $BaTiO_3$, and 3.0 nm for $BiFeO_3$.[31, 32] However, there are several exceptions, for example in the ultra-thin strain-free $SrTiO_3$ film, ferroelectricity can be even enhanced in low dimensions.[33] For recently emerged 2D ferroelectric materials, there are already several comprehensive reviews [34-36]. Here, we only focus on several heavily studied 2D examples and their structural properties in the next section.

### 3. 2D ferroelectric materials

3.1 MXenes. In 2011, Naguib *et al.* stripped $Ti_3C_2$ at room temperature to obtain $Ti_3C_2$ nanostructures, which opened a new area for 2D materials.[37] The general expression of 2D MXenes is $M_{n+1}X_n$, where M = early transition metal, and X = C or N, *n*=1, 2, 3. This type of 2D materials is usually made from a huge family of the MAX phase. [38-40] More than 30 different MXenes have been synthesized and predicted so far.[37] Researchers have found that MXenes has a variety of chemical properties, which makes it widely being used. Among them, energy storage is the most studied application, in addition to water purification[41] catalysis,[42-45], and reinforced composite materials. [46] Since the strength of the M-A bonds and the M-X bonds are different, MXenes are often synthesized by controlling the temperature and using corrosion methods[47]. Due to the different reagents used in the corrosion process, the surfaces of MXenes could have different functional groups, such as O, OH, F, Cl, and S, to modify electronic properties of MXenes. The surfaces of traditional MXenes are often highly reactive, which provide the possibility to induce ferroelectricity. [48, 49] For example, Chandrasekaran *et al.* used first-principle calculations to obtain the polarization value of $Sc_2CO_2$, which is 1.60μC/cm 2.[50] As shown in **Figure 2**, Zhang *et al.* investigated three types of ferroelectric MXene phases (Type I: $Nb_2CS_2$ and $Ta_2CS_2$; Type II: $Sc_2CO_2$ and $Y_2CO_2$; Type III: $Sc_2CS_2$ and $Y_2CS_2$) by using high-throughput search and DFT calculations,[51] and they found that these structures could exhibit robust ferroelectricity in both the out-of-plane and in-plane directions.

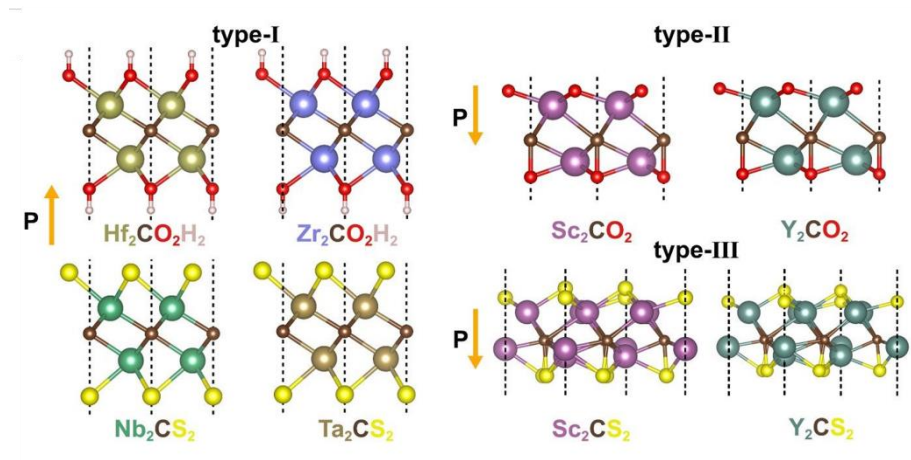

**Figure 2**. Ferroelectric MXenes. The orange arrows indicate the direction of out-of-plane polarization. Reproduced with permission from ref[51]. Copyright 2020 Nanoscale.

3.2 Group IV monosulfur compounds. The general expression of group IV monosulfur compound is MX (M=Sn or Ge, X=Se or S) which have the same orthogonal structure as black phosphorus.[52-55] Recently, it has been shown that these group IV monosulfide monolayers are multiferroic with giant ferroelectric and ferroelastic coupling.[56] As shown in **Figure 3a**, the lone pair electrons of Ge/Sn ions may make their structure distorted, resulting in in-plane switchable ferroelectricity. Moreover, their Curie temperature was predicted to be 326-6400K, which is an unimaginable value in 2D materials.[56] In 2016, Chang *et al.* reported the polarization performance of ultra-thin tin telluride (SnTe) at the liquid helium temperature.[57] Using STM and Scanning Tunneling Spectroscopy (STS), stable fringe regions, lattice distortion, band bending, and electrical polarization operations were observed in SnTe with a thickness of 1 u.c. The polarization characteristic of SnTe is considered to be caused by the change of the ideal square formed by Te atoms into a parallelogram, as shown in **Figure 3b**. Moreover, 2-4 u.c SnTe films also showed strong ferroelectricity at the room temperature.[58] Inspired by the study of 2D ferroelectric SnTe, Wan *et al.* predicted possible ferroelectric phases in group IV tellurides. [58-60]

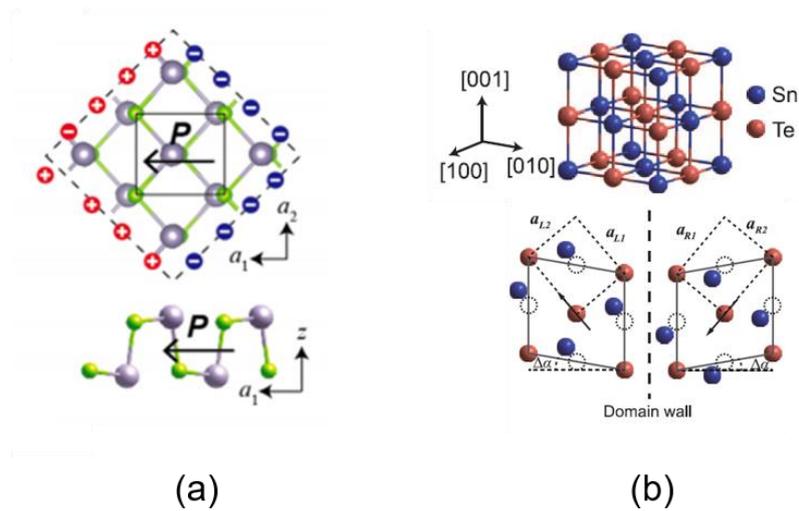

**Figure 3**. (a) Lattice structure of SnSe monolayer. The arrow in the figure indicates the direction of ferroelectric polarization, Reproduced with permission from ref [53]. Copyright 2020 Nano Lett. (b) Schematics of the SnTe crystal structure and the lattice distortion and atom displacement in the ferroelectric phase. Reproduced with permission from ref [57]. Copyright 2016 Science.

3.3 III$_2$-VI$_3$ compounds. These compounds are stable single-layer 2D ferroelectric material based on III-VI compounds, with a general expression being A$_2$B$_3$, where A = Al, Ga, In, B = S, Se, Te. As early as 1990, Abrahams proposed that In$_2$Se$_3$ may be a ferroelectric based on structural analysis.[61] The single-layer structure of In$_2$Se$_3$ consists of Se and In layers alternately arranged by covalent bonds, and the arrangement sequence is Se-In-Se-In-Se. The origin of its ferroelectricity is believed to be the unequal interlayer spacing between the Se layer and two adjacent In layers.[62] In$_2$Se$_3$ usually has five known forms (e.g. α, β, γ, δ and κ), among which the α phase is considered to be the most stable layered structure at the room temperature (**Figure 4**).[63] Zhou et al. reported that the multilayer α-In$_2$Se$_3$ with a film thickness of 10 nm has piezoelectric and ferroelectric properties.[64] They observed ferroelectric domains through piezoelectric response force microscopy (PFM), with obvious phase contrast and domain wall boundaries. Xiao et al. reported the intrinsic ferroelectric properties in thin In$_2$Se$_3$ crystals with an atomic layer thickness of less than 3 nm, and verified the polarization locking mechanism by the second harmonic generation (SHG) spectroscopy measurement and PFM.[65] Ding et al. systematically investigated a series of A$_2$B$_3$ materials by DFT calculations,[63] and the results show that the ferroelectric phases FE-ZB and FE-WZ are also the ground states of Al$_2$S$_3$, Al$_2$Se$_3$, Al$_2$Te$_3$, Ga$_2$S$_3$, Ga$_2$Se$_3$, Ga$_2$Te$_3$, In$_2$S$_3$ and In$_2$Te$_3$. In short, this type of 2D ferroelectric material has inherent out-of-plane and in-plane electric polarization. With the aid of an appropriate out-of-plane or in-plane electric field, these electric polarizations can be converted through a dynamic approach.

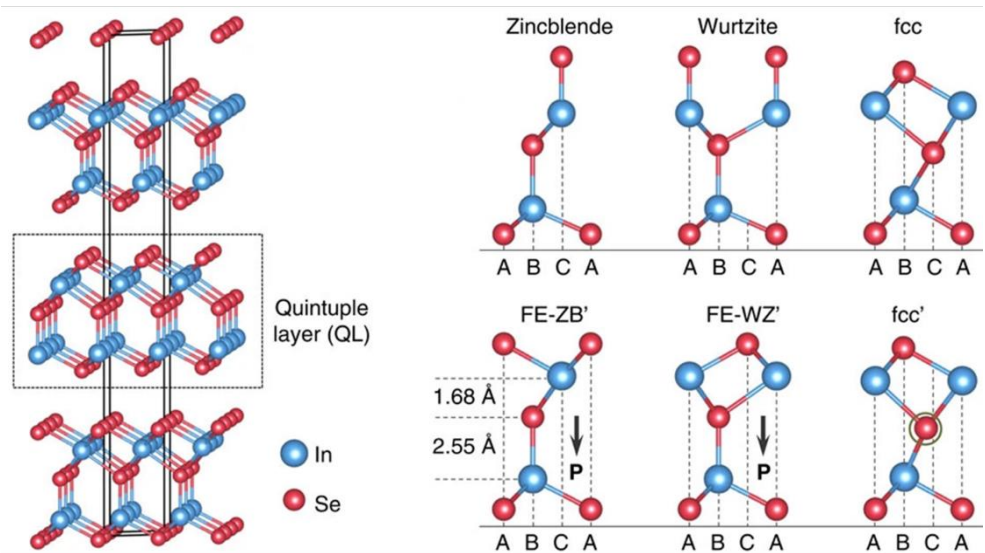

**Figure 4**. Three-dimensional crystal structure of layered $In_2Se_3$ (left), and side views of several representative structures of one QL $In_2Se_3$ (right). Reproduced with permission from ref [63]. Copyright 2017 Nat. Commun.

3.4 Transition metal dichalcogenides. Transition metal disulfides can reduce their atomic structures to the nanometer scale and do not require lattice reconstructions.[66] Their structures are non-centrosymmetric, which provides a basis for the existence of ferroelectricity.[67] In 2014, piezoelectricity was found in $MoS_2$ monolayers, yet this characteristic only exists in odd-numbered layers.[68][69] In the same year, 1T single layer $MoS_2$ (d1T-$MoS_2$) was predicted as a ferroelectric material. The two sulfur lattice planes are arranged in a way that each Mo site becomes the center of inversion, making 1T-$MoS_2$ to be a promising candidate for ferroelectricity (**Figure 5a,b**).[70] Similar calculations were extended to some other 1T $MX_2$ (M = Mo, W; X = S, Se, Te) with twisted octahedral coordination structures ($t$-$MX_2$), as shown in **Figure 5c**.[71] They found that all $t$-$MX_2$ monolayers with d2 metal ions show spontaneous dielectric polarization, which makes TMDs materials promising for 2D ferroelectric materials. In 2018, the study of topological semimetal tungsten ditelluride ($WTe_2$) once again enriched this type of ferroelectrics,[71] and the authors proposed that the polarization of $WTe_2$ is caused by the relative movement of the electron cloud related to the ion nucleus.

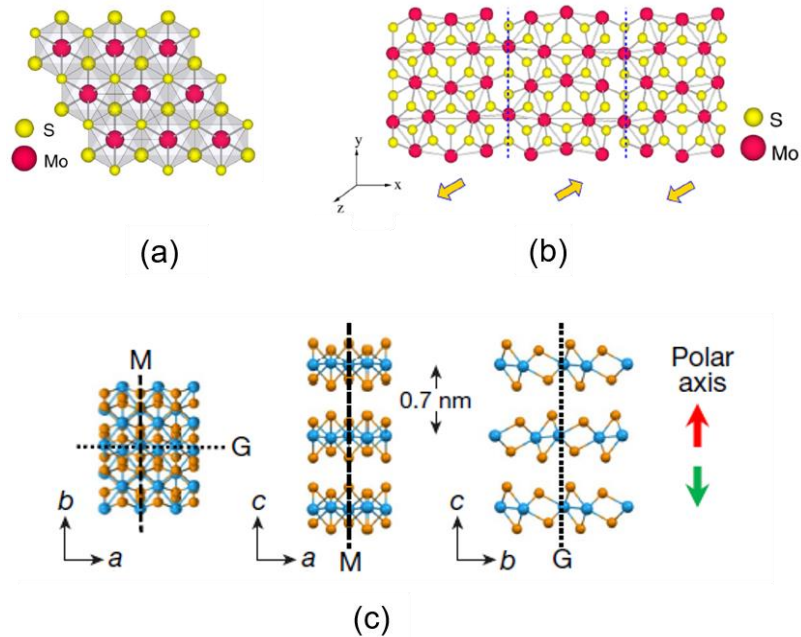

**Figure 5**. (a). Top view of the structure of a monolayer of MoS$_2$ in c1T polytypical form with octahedral coordination of Mo atoms. (b) Structure of the domain wall (blue dashed line) between up and down polarized states of d1T. Reproduced with permission from ref [70], Copyright 2017 Phys Rev Lett. (c) Structure of three-dimensional 1T′ WTe2, showing the mirror plane (M; dashed), glide plane (G; dotted) and polar c axis (red arrow, up; green arrow, down). W atoms are blue; Te atoms are orange. Reproduced with permission from ref [71], Copyright 2018 Nature.

3.5 Transition metal phosphochalcogenides (TMPs). TMPs are a large class of van der Waals layered solids, and their general expression is ABP$_2$X$_6$, where A or B is a combination of monovalent/trivalent or divalent/divalent; and X is chalcogenide, including S, Se, Te, *etc*.[72, 73] One of the important characteristics is that the van der Waals forces between the layers are very weak, hence, a thin film can be easily peeled from the single crystal. The most representative example is the CuInP$_2$S$_6$, which contains a sulfur skeleton, and the triangular patterns of Cu, In and P are filled with octahedral voids (**Figure 6a and 6b**). This special structure makes it possible to have ferroelectric or anti-ferroelectric characteristics.[74] In 1994, the paraelectric-ferroelectric phase transition of CuInP$_2$S$_6$ was observed experimentally at 315 K.[75] Later, Belianinov *et al*. reported that the thickness of the ferroelectric CuInP$_2$S$_6$ film can be achieved as low as 50 nm, which provides an opportunity to obtain ferroelectric CuInP$_2$S$_6$ at nanometer scale. After that, Liu *et al*. reported the room temperature ferroelectricity of ultra-thin CuInP$_2$S$_6$ (~ 4 nm) and the transition temperature is 320K.[76] This spontaneous polarization is caused by the moving of Cu sublattice from center symmetry to the In sublattice. The anti-alignment dipole greatly reduces the depolarization field inside CuInP$_2$S$_6$, so that the stable polarization is shown at an ultra-thin thickness.[77, 78] In 2017,

Xu *et al*. used first principles to design a new TMTP material—AgBiP$_2$Se$_6$, as shown in **Figure 6c**. Among them, the deviation of the Ag$^+$ and Bi$^{3+}$ causes the adjacent unit cells to have different polarizations in the same direction, thus forming a stable ferroelectric phase with a polarization value of about 0.2μC cm$^{-2}$.[79] **Figure 6d** is the band edge of single-layer AgBiP$_2$Se$_6$ calculated by the HSE method, and the results show that the VBM of AgBiP$_2$Se$_6$ is 0.41 eV higher than the oxidation potential of water, while the CBM is 0.67 eV lower than the hydrogen reduction potential. Therefore, the single-layer AgBiP$_2$Se$_6$ exhibits the ability of a water-decomposing catalyst.

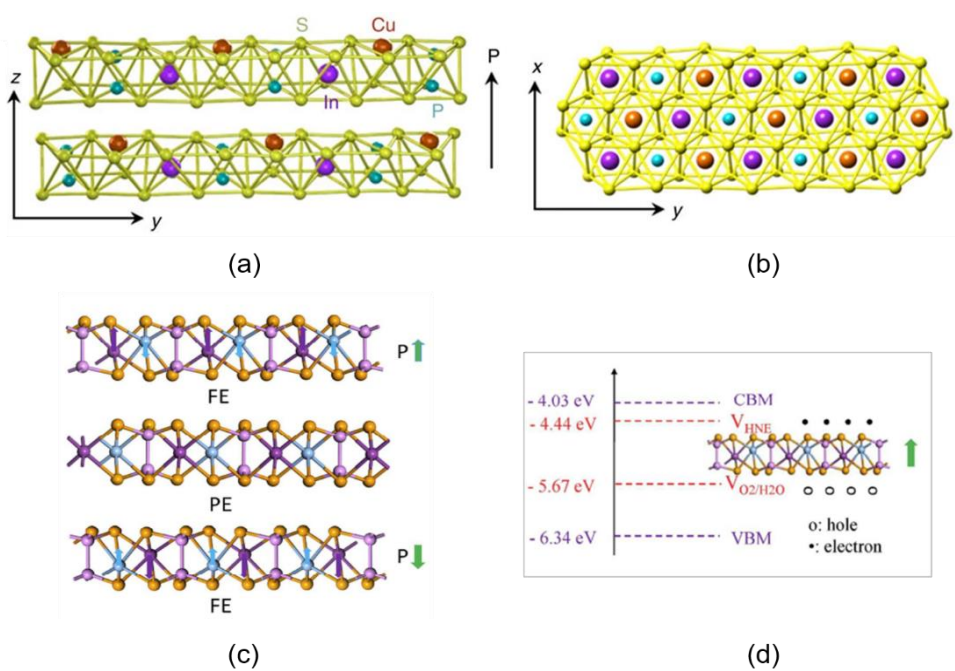

**Figure 6**. The side view (a) and side view (b) for the crystal structure of CuInP$_2$S$_6$ with vdW gap between the layers. The polarization direction is indicated in by the arrow. Reproduced with permission from ref [76]. Copyright 2016 Nat. Commun. (c) Side view of two distorted ferroelectric phases with different polarization directions (top and bottom) and high symmetry paraelectric phase (center). (d) Band edge of single layer AgBiP$_2$Se$_6$. Reproduced with permission from ref[79]. Copyright 2017 Nanoscale.

## 4. Applications of ferroelectric materials in catalysis

Catalysis reaction involves complex reaction processes, which are often affected by several factors. For example: the binding strength between the adsorbate and the reaction surface, the electron transfer between the catalyst and the reactant, and the separation and recombination ability of photo-generated charges should also be considered in photocatalysis. Ferroelectric materials have also been shown to play a variety of promoting roles in the catalytic reaction, which will be summarized in the next sections.

### 4. 1 Gas adsorption

Previous studies have known that ferroelectricity affects physical and chemical properties of the material's surface, which may enhance their interactions with gas molecules.[80] In 1952, Parravano observed that the CO oxidation rate of sodium niobate and potassium niobate near the Curie temperature was abnormal,[81] which opens up a new direction for the application of ferroelectric materials in chemistry. In 2006, Ramos-Moore *et al*. used the temperature programmed desorption method (TPD) to study the $CO_2$ desorption energy of $KNbO_3$ at different temperatures.[82] **Figure 7a and 7b** shows the $CO_2$ desorption curves of $KNbO_3$ powder with three different heating rates. It can be found that there are two peaks in the figure, but there is only one peak in $KTaO_3$. Interestingly, the position of the second peak is somehow independent on the heating rates, which is always near the phase transition temperature. The authors concluded that the appearance of the second desorption peak (580K, after the phase transition temperature of 433K) is related to the ferroelectric phase transition of $KNbO_3$, and $CO_2$ is adsorbed at the edge of the materials. In 2007, Yun *et al*. used the TPD method to study the adsorption properties of $LiNbO_3$ to polar and non-polar molecules (two polar molecules, acetic acid and 2-propanol, and one non-polar molecule, dodecane).[80, 83] The results show that the adsorption and desorption temperature of the polar molecule acetic acid on the positive electrode surface of $LiNbO_3$ is about 101 K higher than that on the negative electrode surface (**Figure. 7c and 7d**), while the non-polar molecule is not affected by the polarized surface. With the same technique of TPD, Zhao *et al*.[84] investigated the polarization-dependent adsorption coefficient of ethanol on $BaTiO_3$, and the results show that ferroelectric materials not only change the number of active sites, but also affect chemical reactions at the surface. Based on the experiments, the authors proposed a "precursor-mediated" mechanism, that is, ethanol first exists on the surface via physical adsorption, and when it encounters a defect site, it may react with surfaces and become chemical adsorption. On the other hand, DFT calculations showed that the chemisorption energies of a series of small molecules on the $Pt(100)/PbTiO_3$ surface could be dramatically changed by reorienting the polarization direction of the substrate[85].

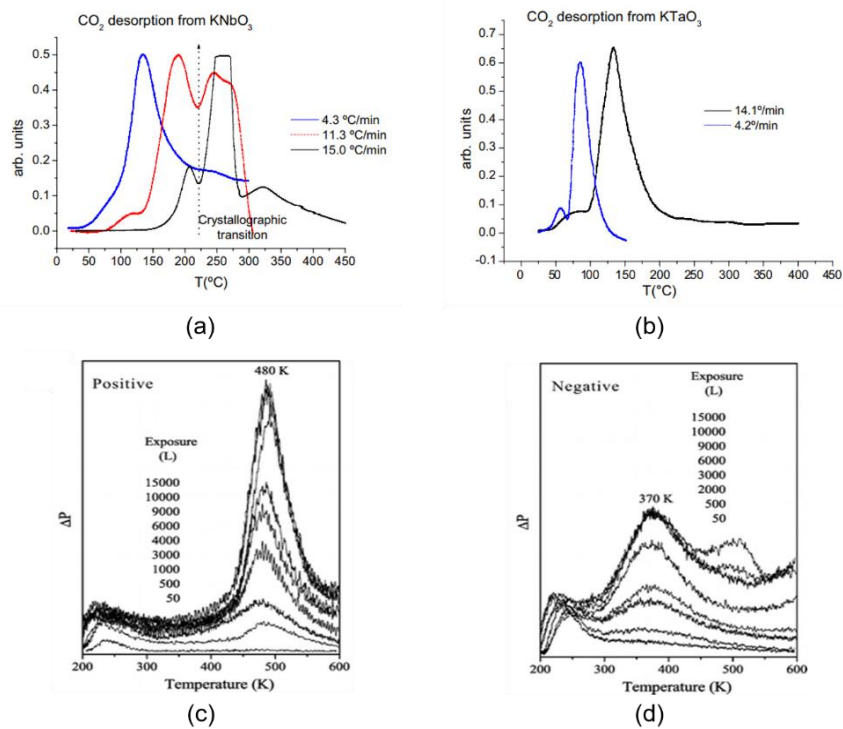

**Figure 7**. (a) Carbon dioxide desorption curves for powder of $KNbO_3$ at three different heating rates. (b) Carbon dioxide desorption curves for powder of $KTaO_3$ at two different heating rates. Reproduced with permission from ref [82]. Copyright 2006 Surface Science. TPD traces for 2-propanol on positively (c) and negatively (d) poled LiNbO3 (0001) surfaces. Reproduced with permission from ref[83]. Copyright 2007 J. Phys. Chem. C.

Based on the DFT calculations, Kakekhani et al.[86] studied the influence of external fields on the chemical properties of the surface of ferroelectric materials, $PbTiO_3$ (PTO), and the authors demonstrated that the surface periodically could alternate between oxidizing, inert and reducing behavior under cyclic polarization conditions[7, 87, 88]. Subsequently, they applied this knowledge to design catalytic cycles for several industrially important reactions. The results show that on the negatively polarized surface of PTO, $SO_2$ will get O atoms from the surface of PTO to generate $SO_3$. The studies of NO, $NO_2$, $N_2$, and $O_2$ on the polar surface shows that polarization could be used as a switch to control their binding energies, and the alternation of positive and negative polarity could even overcome the problem of O inhibition for the direct decomposition of $NO_x$. The only limitation is that $NO_x$ decomposition will not occur on a low PTO cover layer (⩽0.25 Monolayer), which might be achieved via 2D ferroelectric materials.[7] Concerning the direct partial oxidation of $CH_4$ to $CH_3OH$, the authors found that the reaction prefer to occur on the negatively poled oxidizing surface, of which the energy barrier is ca. 1:1 eV, and the binding energy of $CH_3OH$ is almost negligible.

Recently, Tang et al. showed that the reversible polarization transition of the 2D ferroelectric-$In_2Se_3$ monolayer can effectively regulate the adsorption energies, electron transfer and magnetic moment of the adsorbed metal porphyrazine (MPz) molecules by DFT calculations.[89] The electrostatic potential difference between MPz molecules and $In_2Se_3$ monolayers illustrates that the unique behavior of adsorption energies and electron transfer, and the occupancy changes of $d$ orbitals at different polarizations are due to the magnetic ferroelectric tuning. At the same time, their research on the adsorption of a series of gas molecules (e.g. $NH_3$, $NO$, $NO_2$) on $In_2Se_3$ monolayer show that there is an obvious polarization-dependent gas molecule adsorption behavior on the surface of the ferroelectric $In_2Se_3$ monolayer, which can achieve reversible gas adsorption and release controlled by the ferroelectric switch.[90] Later, the same procedure has been employed to control the adsorption behavior of $O_2$, $CO$ and $H_2O$ on the Fe/Mn doped graphene by taking $In_2Se_3$ as the substrate[91]. The above-mentioned phenomenon are basically originated from the synergistic effect of different electrostatic potentials and electrons, which is caused by the band arrangement between the molecular orbitals of the gas front and the edge states of the baseband.[90] As a summary, the controllable ferroelectric adsorption behavior and molecular multiferroic characteristics could be widely used in gas adsorption.[84]

**4.2 Catalysis**

In photocatalytic reactions, polarization can well separate light-induced charge carriers to improve photocatalytic efficiency, e.g. the recombination of photogenerated carriers and the reverse reaction of intermediate species are the main factors that limit the efficiency of water photolysis catalysts.[92] Ferroelectric materials have high surface energy and affect the electrical properties of the generated charge carriers, and the spatial selectivity of its surface may promote photocatalysis. It is known that the mechanism of the spatial selectivity of ferroelectric materials is due to the spontaneous polarization, which can bend energy bands of crystals, thereby causing photogenerated electrons and vacancies to move in opposite directions, promoting the separation of holes and electrons[93]. In fact, a series of studies on the application of ferroelectric materials to photocatalysis have been carried out, as summarized in a review by Kakekhani[87]. Almost 30 years ago, Inoue et al.[94] studied the photo-assisted water splitting on lead zirconate titanate (PZT) ceramics, with focus on the effects of potassium addition on the photoelectric performance of hydrogen production from water. As shown in **Figure 8a**, when $Pb_{1-x}K_xNb_{20}$(PKN) has polarization, it not only shows a large transient pyroelectric current, but also the subsequent current significant decays to almost zero. Recently, 2D ferroelectric materials starts to attract more attention in the vein of photo-assisted water splitting[95]. A representative example was shown by Yang *et al*.[96] in which the authors investigated the potential feasibility of a series of 2D $M_2X_3$ (M = Al, Ga, In; X = S, Se, Te) materials for photocatalytic water splitting by using first-principles calculations (**Figure 8b**). The results show that all candidates are verified to be suitable for water splitting, of which the predicted solar-to-hydrogen efficiency of $Al_2Te_3$, $Ga_2Se_3$, $Ga_2Te_3$, $In_2S_3$, $In_2Se_3$, and $In_2Te_3$ are larger than 10%. In particular, the $In_2Te_3$ shows a supersizing high solar-to-hydrogen efficiency of 32.1%.

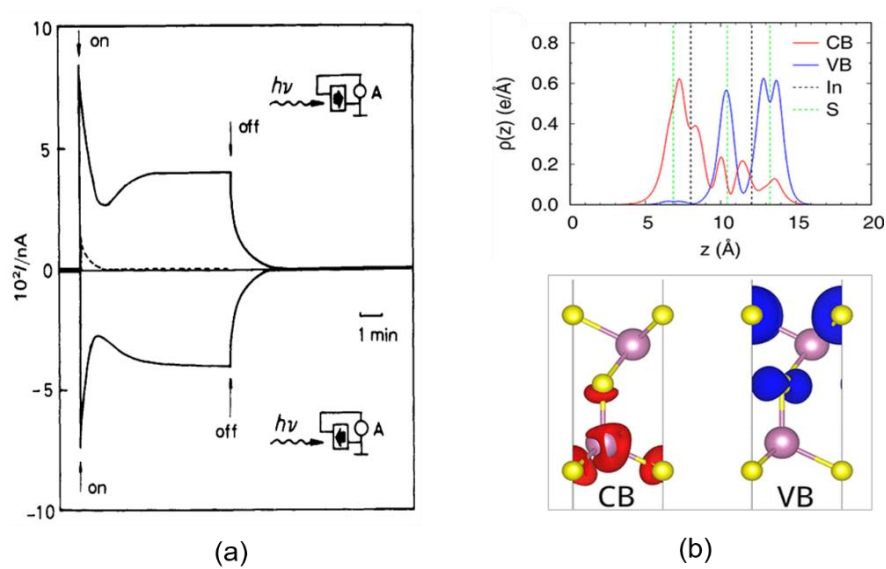

(a)          (b)

**Figure 8.** (a) Photoresponse of PKN ($x$ = 0.006). Reproduced with permission from ref [94]. Copyright 1990 Journal of the Chemical Society, Faraday Transactions. (b) Distributions of VB and CB along the vertical direction (z axis) for $In_2S_3$ monolayers (upper panel), and spatial distribution of CB and VB (lower panel). Reproduced with permission from ref[96]. Copyright 2018 Nano Lett.

Another widely studied application of ferroelectric materials as photocatalysts is the reduction of $Ag^+$ to Ag. In 2001, Jennifer *et al*. irradiated $BaTiO_3$ with ultraviolet light in an aqueous solution containing dissolved $Pb^{2+}$ or $Ag^+$, and found that $Ag^+$ reduction occurred on the positively polarized surface, while more $Pb^{2+}$ oxidation occurred on the negatively polarized surface[97]. Later, Bhardwaj *et al*.[98] studied effects of Sr doping on the catalytic performance of $BaTiO_3$, and found that in $Ba_{1-x}Sr_xTiO_3$, the reduction effects on $Ag^+$ is the largest when $x$=0.26. Subsequently, Schultz *et al*. [99] detailly analyzed the effects of crystal and domain orientation on the photochemical reduction of Ag on $BiFeO_3$. They found that $BiFeO_3$ is beneficial to the reduction of $Ag^+$, and this phenomenon corresponds to the ferroelectric domain structure. Moreover, they calculated the energy band diagrams of bulk $BiFeO_3$, and the results show that when the polarized surface comes into contact with the solution, the energy band will bend.

In electrocatalysis, the interactions between the reactant and the catalyst could be also affected by the polarization.[100] In 2018, Kushwaha et al.[101] investigated the effect of ferroelectric $Bi_{0.5}Na_{0.5}TiO_3$ (BNT) with different polarization values on the electrocatalytic oxygen evolution reaction (OER) (**Figure 9**). They found that iron polarization can be used to adjust the OER activity of BNT, which has a certain relationship with the built-in electric field generated by the polarization in BNT. The reason for this phenomenon is similar to that of photocatalysis, that is, the increase of iron polarization causes the energy band of the negative electrode surface to bend upward, while the energy band of the positive electrode surface bends downward. The later helps to effectively transfer the generated charge from the surface of the catalyst to the electrolyte, thereby promoting the redox reaction. Moreover, Li et al.[102] studied the influence of layered ferroelectric oxides with different Co doping concentrations on the OER reaction. The results show that with the increase of Co content, more oxygen absorbed on the surface. The OER performance of the pure cobalt-based oxide is 100 times higher than that of the pure iron-based oxide, which exceeds the precious metal benchmark $IrO_2$ catalyst. Similarly, iron polarization helps to reduce the overpotential by 70 mV for OER reaction on the $Bi_5CoTi_3O_{15}$ (BCTO) composite catalyst, thereby providing a current density of 10 mA $cm^{-2}$.[103]

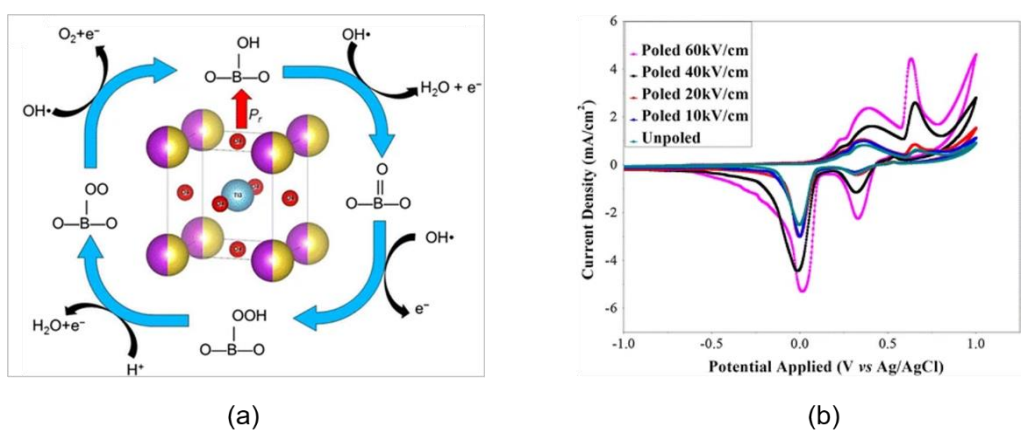

**Figure 9**. (a) Proposed OER mechanism for BNT ferroelectric catalysts. (b) Cyclic voltammograms of poled and unpoled BNT pellet electrodes in 0.1 M KOH at a scan rate of 100 mV/s. Reproduced with permission from ref [101]. Copyright 2017 Journal of Materials Science.

Moreover, the application of ferroelectric materials as electrocatalysts has been applied in a series of small molecules activation. Taken $In_2Se_3$ as an example, Kim[104] constructed a catalytic system with $In_2Se_3$ on top of the hexagonal close-packed (HCP) Co metal slab, and investigated its hydrogen evolution reaction (HER) performance via *ab initio* calculations. The author found that
that the reversible polarization switching of $In_2Se_3$ can turn the HER activity of the heterostructure ON and OFF. Meanwhile, Kan *et al* [105] studied we a Pt single-atom catalyst by taking the ferroelectric monolayer $In_2Se_3$ as the substrate, and the results show that Pt atom strongly interact with $In_2Se_3$, and has a negative clustering energy. Importantly, the Pt/$In_2Se_3$ catalyst shows extremely high activity for CO oxidation, which could be even regulated by the ferroelectric switch of monolayer $In_2Se_3$. Very recently, Kou and his co-works[106] examined a series of catalytic systems, TM@$In_2Se_3$ (TM = Ni, Pd, Rh, Nb, and Re), for the electrochemical reduction of $CO_2$. Through extensive *ab initio* calculations, the author identified the Rh@$In_2Se_3$ catalyst as a highly efficient electrochemical catalyst with a fairly low limiting potential (<0.5 V). Interestingly, they found that Re@$In_2Se_3$ and Nb@$In_2Se_3$ catalysts can realize selective generation of desired products via polarization switching.

**Summary and Outlook**

This review mainly discusses the intrinsic features of traditional ferroelectric materials, 2D ferroelectric materials, and their role in the catalytic reaction process. The polarization characteristics of ferroelectric materials show important influence on the catalytic activities, the electron transfer and the active centers, which could be used to boost the reaction efficiency and control the chemical reactions. Although the researches of ferroelectric materials & phenomena in the chemical catalytic process are limited, great achievements and interesting findings have been made, which provide new ideas for adjusting the catalytic reaction path and controlling the selectivity. Here, we propose several possible research directions of chemical catalysis based on ferroelectric materials as following.

(1) Compared with traditional perovskite oxides, 2D ferroelectrics have obvious advantages like higher stabilities and larger reaction surfaces, which will have more significant effects on the catalytic process. Although a few 2D ferroelectric materials like $In_2Se_3$, $CuInP_2S_6$ and SnTe have been experimentally synthesized, more (up to 100 candidates) are theoretically predicted. Despite the promising potential as we summarized above, the quite small repository of the ferroelectric family severely limited its application in chemical catalysis. The deeper understanding for the ferroelectric switching and its effect on the chemical reaction is urgently required.

(2) Beside the application of physical properties, the chemical applications of ferroelectric materials are an important research field. District from the common catalysts, current research shows that the polarization of ferroelectric materials can effectively control the reaction path due to the reversibility and its effects on the chemical activities. The dynamic surfaces provide the excellent platform to study the switchable chemistry. As a result, the selectivity of products will be greatly improved. Among them, 2D ferroelectric materials which have the advantages of high stability, no surface dangling bonds, and high Curie temperature, are promising candidates to achieve high-efficient catalysis process. Even so, the balance between reactivity and selectivity is still an open question to be addressed. Meanwhile, the roles of polarization and ferroelectricity on separations of photoexcited charge carrier, prevention of the recombination, optical adsorptions and band offsets during the photocatalytic process need to be clarified.

(3) Due to their properties of negligible vapor pressure, high gas molecule solubility, high electric conductivity and a wide potential window[107], ionic liquids (ILs) have been widely used in gas adsorption, catalysis, and materials science[108]. Taken the $CO_2RR$ reaction as an example, the low solubility of $CO_2$ in aqueous electrolytes might hinder the reduction reaction of $CO_2$, and ILs has become an effective solvent to solve this bottleneck issue. As demonstrated by Masel *et al.*[109], the overpotentials of electrochemical reduction of $CO_2$ to CO is significantly decreased to be below 0.2 volt by taking ionic liquids as the electrolyte. Hence, developing a catalytic system consisting of ILs and ferroelectric materials would also be an important direction.